\documentclass[11pt,a4paper]{article}
\usepackage{epsfig,graphicx,amsfonts,amsmath}

\usepackage{cite}
\RequirePackage{mathrsfs}
\usepackage{color}

\newlength{\dinwidth}
\newlength{\dinmargin}
\def\eq#1{{Eq.~(\ref{#1})}}
\newcommand{\Le}{\left(}
\newcommand{\Ra}{\right)}

\newcommand{\beq}{\begin{equation}}
\newcommand{\eeq}{\end{equation}}
\newcommand{\beqar}{\begin{eqnarray}}
\newcommand{\eeqar}{\end{eqnarray}}
\newcommand{\tu}{\textsl{u}}
\newcommand{\tv}{\textsl{v}}

\newcommand{\rom}[1]{\uppercase\expandafter{\romannumeral #1\relax}}

\date{}
\begin{document}

\title {{~}\\
{\Large \bf Negative mass scenario and Schwarzschild spacetime in general relativity  }}
\author{ 
{~}\\
{\large 
S.~Bondarenko$^{(1) }$
}\\[7mm]
{\it\normalsize  $^{(1) }$ Physics Department, Ariel University, Ariel 40700, Israel}\\
}

\maketitle
\thispagestyle{empty}

\begin{abstract}
  In this note we discuss the hypothesis of a presence of the negative mass matter in the Schwarzschild's spacetime
and demonstrate, that the discrete $PT$ transformation of the coordinates and mass inversion in the Schwarzschild's metric solution is equivalent to the inversion of the 
Kruskal-Szekeres coordinates, \cite{Kruskal}, in the full region of the solution's spacetime.  
As the consequence of the result, it is argued that the
whole Schwarzschild spacetime can be described in terms of regions of positive and negative masses interconnected by the discrete transforms. 
\end{abstract}

\section{Introduction}

 The motivation of the negative mass gravitation scenario in the different cosmological  models is very clear. In any scenario, see \cite{Villata, Chardin} for example, the presence of some kind of repulsive gravitation forces in our Universe helps with explanation of the existence of
dark energy and dark matter in the models, see \cite{Petit1,Kofinas,Fluid,Villata1,Hajd} and references therein.
There are few possibilities to introduce the gravitational repulsive force to the scene. In the scenario of \cite{Villata} it is argued that the repulsive 
gravitation is arising between matter and antimatter (in the sense of charge) particles without an introduction of any negative masses. It is also possible
to consider a model where our Universe antimatter has a negative mass, see for example \cite{Chardin} where the negative mass extension of Kerr's
solution is populated by our world antiparticles which possess not positive but negative masses.
Additionally, we can consider
the negative mass particles as the anti-particles of the regular ones with the mass sign of them changed from positive to negative in comparison to 
the mass sign of the particles and antiparticles of our Universe. In this case, the annihilation between the particles of two types of the matter will happen for the particles of any charge. 
Important, that the gravitation properties of this matter is also described by Einstein equations after the discrete symmetry
transformations applied, see  \cite{Villata} or \cite{Chardin} and \cite{Souriau} for the discrete symmetries applications in the case of the quantum and classical systems.

  The presence of this type of negative mass antimatter as well explains the dark energy existence in our Universe and
in this note we discuss the possible window for the presence this type of the matter in the spacetime solution given by the  Schwarzschild's metric.
The consideration is based on the following main statements:
\begin{enumerate} 
\item
An assumption about the full symmetry between isolated behavior of the positive and negative masses matter. 
Namely, at the absence of our ,"normal", matter, the negative masses matter behaves as usual one, i.e. it has the positive inertial mass and
mutual attraction sign in the  Newton's gravitation law in the week field approximation, see\cite{Bondi}. As the consequence of this request, in the given framework we have to reproduce the usual 
Schwarzschild metric solution in the regions with the negative mass present;
\item
The anti-gravity assumption about the interaction of positive and negative masses which means that the interaction of the matter and antimatter is due the repulsion force, i.e. with repulsion in the week field approximation;
\item
The unified description assumption. Both solutions for different masses must be encoded in the same Einstein's equations, similarly, for example, to the particle-antiparticle interpretation of the solutions of QFT equations.
This statement does not mean only the invariance of the equation under the $CPT$ transformation and mass inversion. 
It postulates, that both solutions, of positive and negative masses,
are present in the whole spacetime which is defined in terms of some suitable coordinate system, Kruskal-Szekeres for example for the Schwarzschild spacetime, and they are subject of the same
solution of the general relativity equations. Therefore,  
any transforms of the interests do not pull out the solutions from the spacetime defined by this way. 
\end{enumerate}
The last request clarify that we do not consider any bi-metric theories of the gravity as in \cite{Petit1,Petit2}, but mainly try to follow to the 
ideas of \cite{Villata,Chardin}, with, nevertheless, a different interpretation of what is provide the possible repulsive gravitational forces in our Universe.

 The first two properties of the matter and antimatter can be explained through the application of the
$CPT$ symmetry transformation to the Einstein equations, see \cite{Villata,Chardin}.
It can be understood as a consequence of the physics laws invariance under the transform. Namely, introducing gravitational source and probe masses in the problem, 
the equations of motion of the source masses of different signs will the same whereas the relative
sign in the Newton's gravitation law will be different when it will be applied for the probe and source masses of the different signs.
Correspondingly, the third statement above can be considered as a condition of the self-consistency of  the negative mass hypothesis. So far, there are no any experimental 
facts about a non-correctness of the general relativity results or predictions. Therefore, it is looks natural to check the possibility of the negative masses presence in
any solution of Einstein equations, beginning from the  Schwarzschild's metric solution.  If no contradictions in the construction of the
negative mass scenario in the present framework will be found as it can be considered as some verification of the 
possibility of it's existing, which, of course, does not prove the reality of the negative mass. In any case the only experimental checks of the hypothesis can prove or disprove it.
Therefore, the main goal of the note, is a check of the possibility to reveal the negative mass solution in the spacetime determined by the Schwarzschild solution. 
In the next Section we discuss the Schwarzschild's metric solution in it's transformation in the proposed framework and the last Section is the Conclusion of the note  with the 
representation of the discussion of possible consequences of the hypothesis.

\section{Schwarzschild metric and negative mass spacetime }

 The easiest way to understand the negative mass structure of the Schwarzschild metric is to consider the metric in the Eddington-Finkelstein coordinates.
For the ingoing Eddington-Finkelstein coordinates $(\tv,\,r,\,t)$, which describe the metric structure in \rom{1}-\rom{2} regions
(see the definitions in \cite{Frolov} for example),  we have as usual:
\beq\label{Sec1}
ds^2\,=\,-\Le 1\,-\,\frac{2\,M}{r}\,\Ra\,d\tv^2\,+\,2\,d\tv\,d r\,+\,r^2\,\Le\,d\theta^2\,+\,\sin^2\theta\,d\phi^2\,\Ra\,,
\eeq
 with
\beq\label{Sec2}
\tv\,=\,t\,+\,r^{*}\,=\,t\,+\,r\,+\,2\, M\,\ln\left|\frac{r}{2\,M}\,-\,1\right|\,.
\eeq
Correspondingly, the outgoing Eddington-Finkelstein coordinates $(\tu,\,r,\,t)$ describes the solution in \rom{3}-\rom{4} regions, the metric there has the following form:
\beq\label{Sec3}
ds^2\,=\,-\Le 1\,-\,\frac{2\,M}{r}\,\Ra\,d\tu^2\,-\,2\,d\tu\,d r\,+\,r^2\,\Le\,d\theta^2\,+\,\sin^2\theta\,d\phi^2\,\Ra\,,
\eeq
where
\beq\label{Sec31}
\tu\,=\,t\,-\,r^{*}\,,
\eeq
see \cite{Frolov}.
Now we would like to continue the metric in the region of negative masses performing a substitution $M\,\rightarrow\,-\,\tilde{M}\,,\,\,\,\tilde{M}\,>\,0$\,.
Nevertheless, this naive continuation is not correct, the particles of negative masses must preserving the same  dynamics as the "normal"
ones. Therefore, the time and radial coordinate must be be reversed as well in order to preserve the horizon properties of 
Schwarzschild's black hole solution: $t\,\rightarrow\,-\,\tilde{t}\,,\,\,\,\,\tilde{t}\,>\,0$\, and 
$r\,\rightarrow\,-\,\tilde{r}\,,\,\,\,\,\tilde{r}\,>\,0$\,, see \cite{Villata,Chardin}. The new variables cover the same $R^{1}$ region as the "normal" ones, 
and now we can construct the full spherically symmetrical vacuum spacetime in Kruskal-Szekeres coordinates accurately performing the discrete transforms.
Consider first the $U\,<\,0\,,V\,>\,0$ region (the region \rom{1} with $r\,>\,2M$) of the $(U,V)$ coordinate plane.
We have there:
\beq\label{Sec4}
U\,=\,-\,e^{-\tu / 4M}\,,\,\,\,\,V\,=\,e^{\tv / 4M}\,.
\eeq
Now, in order to continue this quadrant of the space in the region of negative mass we 
perform the discrete transforms. For the incoming coordinate we obtain:
\beq\label{Sec5}
\tv\,=\,t\,+\,r^{*}\,\rightarrow\,\tilde{\tv}\,=\,-\,\tilde{t}\,-\,\tilde{r}\,+\,2\,M\,\Le\,\ln\Le\frac{\tilde{r}}{2\,M}\,+\,1\Ra\,+\,\imath\,\pi\,\Ra\,,
\eeq
after the subsequent $t$ and $r$ reverse and
\beq\label{Sec6}
\tilde{\tv}\,=\,-\,\tilde{t}\,-\,\tilde{r}\,-\,2\,\tilde{M}\,\Le\,\ln\Le\frac{\tilde{r}}{2\,\tilde{M}}\,-\,1\Ra\,+\,2\,\imath\,\pi\,\Ra\,=\,
-\tilde{t}\,-\,\tilde{r}^{*}\,-\,4\,\imath\,\pi\,\tilde{M}\,=\,-\tv\,-\,4\,\imath\,\pi\,\tilde{M}\,,
\eeq
after the final reversion of the mass. For the outgoing coordinate we will have similarly:
\beq\label{Sec7}
\tu\,=\,t\,-\,r^{*}\,\rightarrow\,\tilde{\tu}\,=\,
-\,\Le\,\tilde{t}\,-\,\tilde{r}^{*}\,\Ra\,+\,4\,\imath\,\pi\,\tilde{M}\,=\,-\,\tu\,+\,4\,\imath\,\pi\,\tilde{M}\,.
\eeq
Therefore, the continuation of \eq{Sec4} coordinates in the region of negative mass will have the following form:
\beqar\label{Sec8}
U\,=\,-\,e^{-\tu / 4M}\,\rightarrow\,\tilde{U}\,=\,e^{-\tilde{\tu} / 4\tilde{M}}\,=\,-\,U\,,\\
V\,=\,e^{\tv / 4M}\,\rightarrow\,\tilde{V}\,=\,-\,e^{-\tilde{\tv} / 4\tilde{M}}\,=\,-\,V\,.
\eeqar
This inversion of the signs of the $(U,V)$ coordinate axes will hold correspondingly in the all regions of $(U,V)$ plane after the \eq{Sec6}-\eq{Sec7}
analytical continuation.   
Formally, from the point of view of the discrete transform performed in $(U,V)$ plane, the transformations \eq{Sec8}
are equivalent to the full reversion of the Kruskal-Szekeres "time" 
\beq\label{Sec9}
T\,=\,\frac{1}{2}\,\Le\,V\,+\,U\,\Ra\,\rightarrow\,-\,T
\eeq
and "coordinate"
\beq\label{Sec10}
R\,=\,\frac{1}{2}\,\Le\,V\,-\,U\,\Ra\,\rightarrow\,-\,R
\eeq
of the complete Schwarzschild spacetime. The spacetime with 
 $(\tilde{U}\,,\tilde{V})$ variables, obtained after the applied discrete transforms, must be understood as spacetime
with the "watching point" in the regions of the negative mass. Therefore, 
the region \rom{1}  in the $(\tilde{U}\,,\tilde{V})$ plane corresponds to the region \rom{3} in $(U,V)$ plane,
the "white hole" region, \rom{4} in the space of normal mass, will transform to the
 black hole region, \rom{2}, in the  $(\tilde{U}\,,\tilde{V})$  coordinate plane and vise verse. 
Important, that both spaces of normal and negative masses are encoded in the same solution of the Einstein
equation in correspondence to our third proposition.

\section{Conclusion}

 In this short note we discuss the possibility of the presence of the negative mass matter in the world described by Schwarzschild spacetime. Basing on the three statements above,
see Introduction's chapter,
we obtained that the regions \rom{3}-\rom{4} of the Schwarzschild's metric solution in Kruskal-Szekeres coordinates,  disconnected from our "normal" world, 
can be interpreted as the regions populated by the matter of the negative mass.
This result is clarified by the demonstration that the application of the inversion of the mass, radial coordinate and time reversing
in the Schwarzschild's metric solution is equivalent to the full inverse of the  Kruskal-Szekeres coordinates in whole spacetime determined by the
Schwarzschild's metric, 
that is the mostly important result of the present note. In our future research we plan to investigate, therefore, if this result can be justified in the case of more 
complex symmetrical solutions of Einstein equations as well, see \cite{Frolov,Carter}.

 The first important consequence of the obtained result is that the regions of the negative mass as disconnected from the spacetime
of "normal" mass, see for example \cite{Fuller}. Namely, with the
Schwarzschild's metric as the solution of the Einstein equations, 
there is no any direct connection between the regions populated by the negative and "normal" masses. Nevertheless, in this scenario, the white
holes in our part of the spacetime are inverted into the black holes which consists of the mass of another type, this is the only sign of the negative masses in our part of 
the full spacetime
for the case of Schwarzschild's solution, see also \cite{Carter}. 
The influence of these "negative" black holes on the "normal" matter is in the repulsion of the matter around, no other signals can be detected from the objects.
Taking into account, that the negative mass scenario is proposed to explain the dark energy presence in our universe, 
we conclude that in the given framework the
dark energy can corresponds to the mass of the "negative" black holes in the negative mass region of the spacetime. 
Another interesting observation is a possibility of modifications and extensions of some non-canonical theories of cosmology 
in the spacetime with negative mass present, see \cite{Saharov,Linde}.

 Concluding we underline that any proposed scenario of the negative mass existing have to explain the features of our region of spacetime and any cosmological 
properties of the observed universe without redundant modification of existing theories and 
it must be verified in both theoretical and experimental ways.
To our opinion, the further work in this direction it is a very intriguing and interesting  subject of research.  

The author is grateful to M.Zubkov, M.Schiffer and A.Kashi for the numerous and useful discussions.


\newpage
\begin{thebibliography}{99}

\bibitem{Kruskal}	
	M. D. Kruskal, Phys. Rev. 119, 1743 (1960);
G. Szekeres, Publ. Mat. Debrecen 7, 285.
(1960).

\bibitem{Villata} 
  M.~Villata,
  EPL {\bf 94}, no. 2, 20001 (2011);
  Annalen Phys.\  {\bf 527}, 507 (2015).

\bibitem{Chardin}	
G.~Chardin,
  Hyperfine Interact.\  {\bf 109}, no. 1-4, 83 (1997).	
	R.~J.~Nemiroff, R.~Joshi and B.~R.~Patla,
  JCAP {\bf 1506}, 006 (2015);
G.~Kofinas and V.~Zarikas,
  Phys.\ Rev.\ D {\bf 97}, no. 12, 123542 (2018);
  G.~Manfredi, J.~L.~Rouet, B.~Miller and G.~Chardin,
  Phys.\ Rev.\ D {\bf 98}, 023514 (2018).


\bibitem{Petit1} 
  J.-P.~Petit,
  Astrophys.\ Space Sci.\  {\bf 226}, 273 (1995).

\bibitem{Kofinas}
A.~A.~Baranov,
  Izv.\ Vuz.\ Fiz.\  {\bf 11}, 118 (1971);
	A.~D.~Dolgov,
  arXiv:1206.3725 [astro-ph.CO].
	
	
	\bibitem{Fluid}
	J.~S.~Farnes,
  A\&A 620, A92 (2018).

  \bibitem{Villata1}
  M.~Villata,
  Astrophys.\ Space Sci.\  {\bf 339}, 7 (2012);
  Astrophys.\ Space Sci.\  {\bf 345}, 1 (2013).
	
	\bibitem{Hajd}
	D.~S.~Hajdukovic,
  Astrophys.\ Space Sci.\  {\bf 339}, 1 (2012);
  Phys.\ Dark Univ.\  {\bf 3}, 34 (2014).

	
\bibitem{Souriau}	
J.-M. Souriau, "Structure of dynamical systems", Progress in Mathematics vol. 149, Springer Science, 1997. 	

\bibitem{Petit2}
J.~P.~Petit and G.~d'Agostini,
  Astrophys.\ Space Sci.\  {\bf 354}, no. 2, 2106 (2014);
  Mod.\ Phys.\ Lett.\ A {\bf 29}, no. 34, 1450182 (2014);
  Astrophys.\ Space Sci.\  {\bf 357}, no. 1, 67 (2015).
	
\bibitem{Bondi} 
  H.~Bondi,
  Rev.\ Mod.\ Phys.\  {\bf 29}, 423 (1957).	
	
\bibitem{Frolov}	
S.Chandrasekhar, "The mathematical theory of black holes", Clarendon Press Oxford, 1983;
V.P.Frolov and I.D.Novikov, "Black holes physics", Kluwer Academic Publishers, 1998.	

\bibitem{Carter} 
  B.~Carter,
Phys.\ Rev.\  {\bf 141}, 1242 (1966);
 Phys.\ Rev.\  {\bf 174}, 1559 (1968).

\bibitem{Fuller} 
  R.~W.~Fuller and J.~A.~Wheeler,
  Phys.\ Rev.\  {\bf 128}, 919 (1962).
	
\bibitem{Saharov}	
A.D.Sakharov, JETP Lett. 5, 24 (1967); JETP 49, 594 (1979); JETP 52, 349 (1980).	

\bibitem{Linde}
A.~D.~Linde,
  Phys.\ Lett.\ B {\bf 200}, 272 (1988).
	
	
\end{thebibliography}
\end{document}